# Photoexcited Small Polaron Formation in Goethite (α-FeOOH) Nanorods Probed by Transient Extreme Ultraviolet Spectroscopy

*Ilana J. Porter[†,‡], Scott K. Cushing[†,‡], Lucas M. Carneiro[†,‡], Angela Lee[†], Justin C. Ondry[†,§], Jakob C. Dahl[†,§], Hung-Tzu Chang[†], A. Paul Alivisatos[†,§,⊥,∥], Stephen R. Leone[†,‡,∇,](mailto:)\**

[†] Department of Chemistry, University of California, Berkeley, California 94720, United States

[‡] Chemical Sciences Division, Lawrence Berkeley National Laboratory, Berkeley, California 94720, United States

[§] Materials Sciences Division, Lawrence Berkeley National Laboratory, Berkeley, California 94720, United States

[⊥] Department of Materials Science and Engineering, University of California, Berkeley, California 94720, United States

[∥] Kavli Energy NanoScience Institute, Berkeley, California 94720, United States

[∇] Department of Physics, University of California, Berkeley, California 94720, United States

AUTHOR INFORMATION

**Corresponding Author**

*e-mail: srl@berkeley.edu

ABSTRACT Small polaron formation limits the mobility and lifetimes of photoexcited carriers in metal oxides. As the ligand field strength increases, the carrier mobility decreases, but the effect on the photoexcited small polaron formation is still unknown. Extreme ultraviolet transient absorption spectroscopy is employed to measure small polaron formation rates and probabilities in goethite (α-FeOOH) crystalline nanorods at pump photon energies from 2.2 to 3.1 eV. The measured polaron formation time increases with excitation photon energy from 70 ± 10 fs at 2.2 eV to 350 ± 30 fs at 2.6 eV, whereas the polaron formation probability (85 ± 10%) remains constant. By comparison to hematite (α-$Fe_2O_3$), an oxide analog, the role of ligand composition and metal center density in small polaron formation time is discussed. This work suggests that incorporating small changes in ligands and crystal structure could enable the control of photoexcited small polaron formation in metal oxides.

**TOC GRAPHICS**

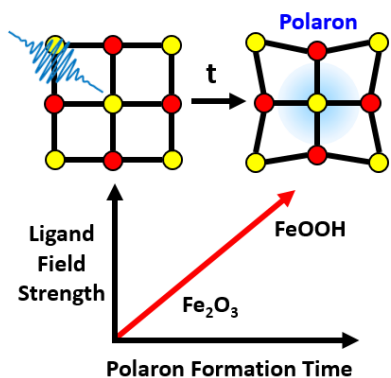

The formation of small polarons in transition metal oxides limits carrier diffusion.[1-6] A small polaron is formed when the electric field of an excess carrier interacts with an optical phonon in a polar lattice, distorting the lattice and trapping the carrier in a local potential well.[7-9] For example, in the iron oxides and oxide hydroxides, small polarons form when electrons self-trap onto an iron center, forcing conduction to occur via phonon-mediated hops between centers.[4-6] In the photoexcited state, it has recently been shown that small polarons form on a sub-100 fs timescale in hematite (α-$Fe_2O_3$).[10] Additionally, small polarons are found to form at the hematite surface in approximately 660 fs.[11] Small polaron formation therefore may control the trapping and lifetime of photoexcited carriers as well as the mobility.

The existence of small polarons is intrinsic to a material since it is governed by the polarity of the lattice.[7] The small polaron formation energy and hopping activation energy, however, are sensitive to the ligand field strength and hopping center density. For example, a linear relationship has been found between the polaron hopping activation barrier and the ionic polarizability at interfaces.[12] In other words, even if small polarons cannot be eliminated in a material, the small-polaron-limited mobility may be controlled through the electronic and structural properties of the material. For example, while goethite (α-FeOOH) and hematite both have an octahedral coordination geometry of oxygen ligands about an $Fe^{3+}$ center, the replacement of some $O^{2-}$ ligands with $OH^-$ ligands in goethite increases the electron density about the Fe-O bonds, creating stiffer, less distortable bonds with higher vibrational frequencies.[13-18] In goethite, the iron atoms fill 1/2 of the interstitial spaces in the hexagonal close-packed array of oxygens, while in hematite the irons fill 2/3 of the interstitial spaces.[19-22] This change in iron center density corresponds to an increase in the Fe-Fe distance by greater than 5% in goethite with respect to hematite.[6,18] These changes in structure and bonding have been shown to increase the polaron hopping activation energies in goethite, decreasing the ground state carrier mobility compared to hematite.[4-6,23] It has yet to be experimentally confirmed whether the same changes to structure and bonding also modulate the excited state small polaron localization and thus the lifetime of photoexcited carriers.

In this study, we measure the polaron formation kinetics of goethite (α-FeOOH) crystalline nanorods using extreme ultraviolet (XUV) transient absorption spectroscopy at the Fe $M_{2,3}$ edge. This pump-probe technique is sensitive to changes in the Fe oxidation state, allowing for the observation of small polaron formation via a signature spectral shift. The small polaron formation time increases with excitation energy from 70 ± 10 fs at 2.2 eV to 350 ± 30 fs at 2.6 eV. Excited electrons are measured to have an 85 ± 10% probability of forming small polarons, and the signal associated with the polaron persists for over 300 ps. Comparison of these trends with hematite, in particular the polaron formation time (180 ± 30 fs average time in goethite nanorods and 90 ± 5 fs in hematite thin films), suggests that polaron formation may be tuned by altering the ligand composition and density of the iron centers, although the role of sample morphology still needs to be investigated.

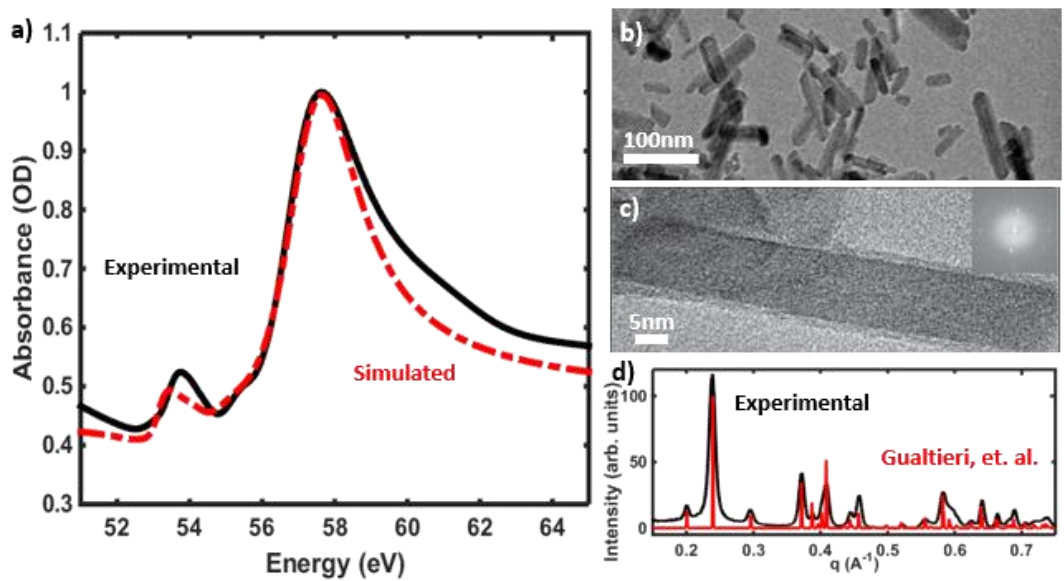

**Figure 1.** a) The ground state XUV absorption spectrum of goethite nanorods (black) and the spectrum simulated using the CTM4XAS software (red). Input parameters for the charge transfer multiplet calculation are summarized in the text. b) A TEM image of the goethite nanorod distribution, confirming the size and the rod-like shape of the particles. C) An HRTEM image of a single nanorod reveals that the entire rod is a single crystal. d) Powder XRD of the sample (black) compared to the stick spectrum of goethite from Gualtieri et al.[19] (red), which confirms that the sample is in the goethite phase.

XUV transient absorption spectroscopy utilizes a visible or near-IR pump and a broadband XUV probe to measure semicore-to-valence transitions, which are sensitive to the oxidation state and bonding environment of first row transition metals. The apparatus, described previously,[24] utilizes the process of high harmonic generation to produce the XUV probe pulses, and it can measure thin solid state samples, such as thin films and nanoparticles, which are suspended on silicon nitride windows.

The ground state XUV absorption spectrum of the Fe $M_{2,3}$ edge for the goethite nanorods is shown in Figure 1a. The observed spectral features, shown in black, are caused by the multiplet splitting between the ground state ($3p^6 3d^5$) and the core hole excited state ($3p^5 3d^6$) and by the ligand field. The simulated spectrum, shown in dotted red, is predicted using a charge transfer multiplet calculation with a value for the crystal field splitting 10Dq of 1.55 eV.[10,24,25] Details of this calculation, including all other parameters, are given in the Supporting Information. The 1.55 eV value is obtained by performing a global fit on the experimental data with the 10Dq value as the fit parameter, resulting in a fit error of 0.01 eV. This crystal field splitting value differs from the visible light fitted value of 1.95 eV,[13,26] with the discrepancy attributed to the core-hole altered crystal field strength of the final state in the x-ray transition.[14,27] Transmission electron microscopy (Figure 1b and 1c) confirms the size distribution and single-crystalline nature of the rods. Powder x-ray diffraction (Figure 1d) is compared to the stick spectra of several common polymorphs of iron oxide and iron oxide hydroxide (goethite in red, hematite and magnetite not shown) to verify the sample identity.

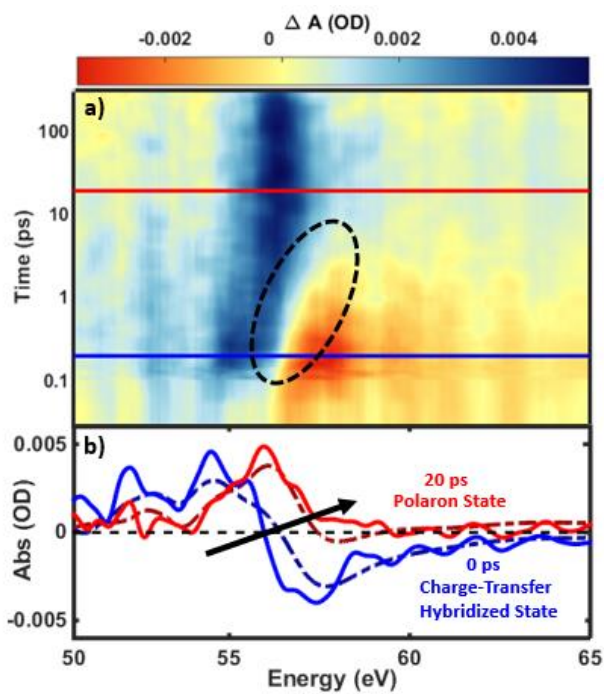

**Figure 2.** a) The transient differential absorption of goethite for the first 300 ps after optical excitation, with a logarithmic time axis. The time axis is offset by 100 fs to improve the clarity of the plot. The solid lines (red and blue) indicate the times for the lineouts shown in panel b. b) The differential absorption at the times indicated in panel a (solid lines) are plotted with the predicted differential absorption spectra for those states (dotted lines). The differential absorption immediately following optical excitation (delay of 0 ps, shown at 0.1 ps in panel a due to the 100 fs offset) matches the prediction for a charge-transfer hybridized state (blue), and the differential absorption at 20 ps matches the prediction for a polaron (red). Solid lines are obtained by averaging the nearest 6 time delays. Details of how the predicted (dotted) spectra are generated is included in the Supporting Information. The arrow indicates the shift of the zero-crossing from approximately 56 eV to 57.5 – 59 eV, which is the most noticeable spectral feature of the polaron formation.

The differential absorption after photoexcitation of the goethite nanorods with 3.1 eV light is shown in Figure 2a. The change in the valence charge density upon photoexcitation alters the multiplet splitting between the 3p core levels and valence levels, modifying the x-ray absorption compared to the ground state. First, when an interband transition is photoexcited in an iron oxide, an electron is transferred from majority O 2p hybridized orbitals to majority Fe 3d hybridized orbitals within 30 fs.[10,24] This charge-transfer hybridized state appears in the differential absorption spectrum as an increase in absorption (blue) between 53 eV and 56 eV and a decrease in absorption (red) between 56 eV and 59 eV, crossing the zero at 56 eV as shown in Figure 2b as a solid blue line. The charge transfer hybridized state is modeled by setting the final state of the absorption to be $Fe^{2+}$ in the charge transfer multiplet simulation (dotted blue line Figure 2b).

Next, the photoexcited carriers thermalize via optical phonon emission. During the electron-phonon scattering process, the optical phonon and electron can couple to form a small polaron. The small polaron can cause an anisotropic lattice expansion, resulting in a splitting of the Fe 3p level.[9] This appears in the differential absorption as a broad increase in absorption between 54 eV and 58 eV, shown in Figure 2b as a solid red line. The polaron differential absorption is modeled as a splitting of the ground state absorption following Carneiro et al.,[10] which is shown as a dotted red line. Briefly, this is accomplished by convolving the ground state spectrum with three delta functions, and further details can be found in the Supporting Information. The evolution from the charge-transfer hybridized state to the polaron state is noticeable by the zero-crossing shift from 56 eV to 57.5 - 59 eV illustrated with the arrow in Figure 2b.

To further understand the small polaron formation dynamics in goethite, transient differential absorption spectra were measured at four visible pump wavelengths spanning 2.2 eV to 3.1 eV. All four spectra are included in Supporting Figure S1. A multivariate regression was performed to decompose the differential absorption spectra into the charge-transfer hybridized state (blue), taken at t = 0 ps, and the polaronic state (red), taken at t = 20 ps. The polaronic state was chosen to be 20 ps in order to minimize the error of the regression, even though the polaron population does not increase after approximately 5 ps. The results of the multivariate regression are shown in Figure 3a for an excitation energy of 3.1 eV, and in the Supporting Information (Figures S2 and S3) for the other excitation wavelengths.

The resulting amplitudes are then fit with a kinetic model representing polaron formation[10] before 20 ps and a stretched exponential representing polaron hopping[28] after 20 ps, as indicated in Figure 3a. The kinetic model for polaron formation is based on a two-temperature rate equation for the hot electron and hot phonon populations, in which an electron and phonon can combine via bimolecular kinetics to create a small polaron. Briefly, the model fits two rate constants, the electron-phonon scattering and the small polaron formation, and two amplitudes, the average hot electron population and average polaron population. The polaron formation probability shown below is the ratio of these population amplitudes. Further details of this model can be found in the Supporting Information.

The polaron formation time and probability resulting from this fit are shown in Figure 3b and 3c, respectively. The polaron formation time has an average value of 180 ± 30 fs across an energy range of 2.2 eV to 2.6 eV and shows a significant increasing trend with pump photon energy up to 2.6 eV, then decreases at 3.1 eV. The polaron formation probability is 85 ± 10% on average across all pump photon energies. Within the experimental variance the probability exhibits a slightly increasing trend with increasing excitation energy. For all excitation wavelengths, the polaron state lives longer than the 300 ps time delay of the measurement, with an average fitted lifetime of 800 ns ± 5 μs (Figure S4). This value is unrealistic for the timescale of the measurement and the standard error of the fit is nonphysical, indicating that the stretched exponential fit cannot be trusted to analyze the polaron decay lifetime beyond the condition that the polaron lives longer than 300 ps.

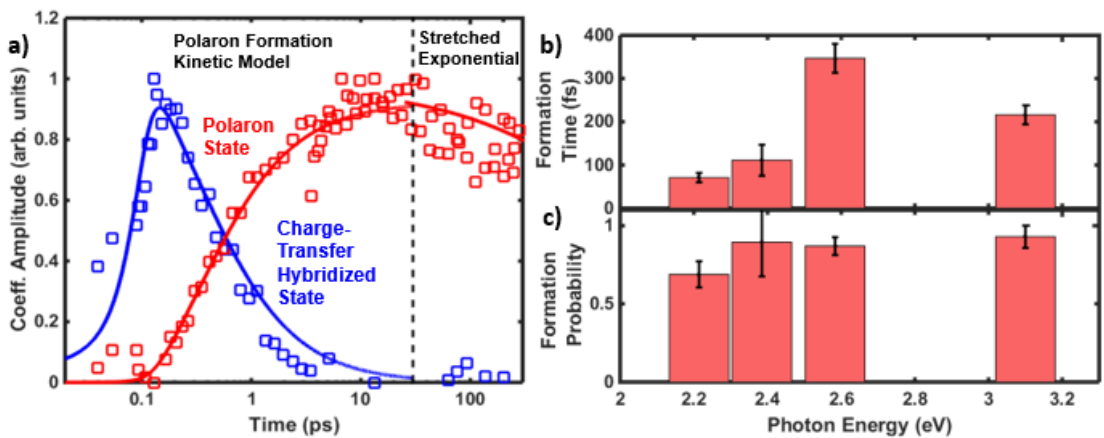

**Figure 3.** a) The result of multivariate regression on the transient absorption spectra of Figure 2a, with amplitudes of the charge-transfer hybridized state and polaron state shown as squares, and model fits shown as solid lines. The dotted black line indicates the split between data that are fit using the polaron formation kinetic model and the stretched exponential polaron decay model. Results here are shown with a logarithmic time axis. The results of the polaron formation kinetic model fits at various pump wavelengths are shown in b) for polaron formation time and c) for polaron formation probability. Error bars shown indicate the standard error.

The small polaron kinetics measured here can be compared to previous measurements of nanocrystalline hematite thin films and to surface-sensitive measurements of polycrystalline and crystalline hematite. While the goethite nanorods exhibit a measured average polaron formation time of 180 ± 30 fs across an energy range of 2.2 eV to 2.6 eV, Carneiro et al.[10] report an average polaron formation time of 90 ± 5 fs for the hematite thin films, which is calculated for the same energy range and using the same kinetic model. This difference in polaron formation time can be first considered in terms of the chemical structure of the two materials. According to a basic kinetic theory for polaron formation, the formation rate should depend on both the attempt frequency and the energy barrier to formation,[8] $\Gamma = \omega e^{-E_{act}/RT}$. Here, $\Gamma$ is the polaron formation rate, $\omega$ is the attempt frequency, $E_{act}$ is the activation barrier, T is the lattice temperature, and R is the Boltzmann factor. For polaron formation, since an electron and an optical phonon must interact, the attempt frequency can be estimated by the LO phonon frequency $\omega_{LO}$. The LO phonon mode with the highest energy will have the fastest scattering rate, so the highest energy $E_u$ mode is chosen for hematite and the highest energy Fe-O $B_{3u}$ mode is chosen for goethite, which have frequencies of 50 fs and 53 fs, respectively.[15-17] The energy barrier for excited state polaron formation can be estimated from the activation energy for electron hopping between the Fe centers $\Delta E_{hop}$, which is 190 meV for hematite and 235 meV for goethite.[4-6] The excited state formation kinetics can be approximated at a lattice temperature of 600 K as previously done for hematite at a similar excitation power density as used here.[8,10]

The ratio of formation times can be estimated by

$$\frac{\Gamma_g}{\Gamma_h} = \frac{\omega_{LO}^g}{\omega_{LO}^h} e^{-(\Delta E_{hop}^g - \Delta E_{hop}^h)/RT} \qquad (1)$$

yielding a value of 42% ± 9%. This predicted ratio matches the average experimental ratio of 50% ± 9% for the polaron formation time measured here for goethite versus that from Carneiro et al.[10] for hematite. Although a simplified estimate, Equation (1) suggests that the change in polaron formation time between hematite and goethite can be accounted for by the difference in electron hopping activation energy between the Fe centers. Since the hopping activation energy is related to the ground state mobility via Marcus Theory,[8] this means that the trends in ground state mobility and polaron formation time may be influenced by similar changes to the ligand field and metal center density.

The above comparison of average formation times neglects the excitation energy dependence of the activation barrier.[8] The average formation time ratio of approximately 50% differs significantly from the ratio of approximately 140% observed at 2.2 eV excitation, of approximately 80% observed at 2.4 eV excitation, of approximately 25% observed at 2.6 eV excitation, and of approximately 45% observed at 3.1 eV excitation. However, the fit amplitudes that relate the initial electron population to the final polaron formation (Figure 3c) are relatively constant over the same excitation energy range. This indicates that multi-phonon effects, which would change the ratio between photoexcited electrons and formed polarons, are not prevalent.

Additionally, Husek et al.[11] report a much longer polaron formation time in hematite surfaces ($640 \pm 20$ fs for polycrystalline, $680 \pm 30$ fs for single crystal) obtained for bulk samples with an XUV probe at near grazing angle. Although the fitting routine and kinetic model differs from the model described above and used by Carneiro et al., a fit of the goethite nanorod data at the same excitation energy and with the model from Husek et al. reveals a similar polaron formation time as described above ($160 \pm 25$ fs at 3.1 eV excitation). This indicates that the much longer polaron formation times measured at the surface could be due to the differences between localization to a 2D surface and localization to a 3D bulk site, precluding a direct comparison of the results.

The trend in polaron formation times with increasing excitation energy is measured to be reversed between hematite (decreasing) and goethite (increasing). This difference could result from the changes in the ligands and the distances between the iron centers, or it may result from the sample morphology. Specifically, the hematite transient spectra from Carneiro et al.[10] were measured for nanocrystalline films, whereas the goethite samples measured here are monocrystalline nanorods. As the excitation energy is increased, the phonon bath must dissipate more heat. For the thin film, excess heat can be dissipated throughout the film and away from the localized excitation spot. This is not the case for the nanorods, as the excess heat cannot be dissipated spatially. The localized non-thermal phonon bath in the nanorods can lead to an increase in polaron hopping and polaron de-trapping, similar to an increase in the sample temperature, possibly explaining the increased formation times at higher excitation energies. However, increased crystallinity could also explain the increased polaron lifetime in the goethite nanorods, as fewer trap-states may be present at which excited carriers can become localized.[11,29-30] A comparison of hematite nanorods to the goethite nanorods is therefore necessary before the change in polaron formation kinetics can be completely attributed to coordination or morphology effects.

The photoexcited polaron formation kinetics of goethite nanorods has been explored with XUV transient absorption spectroscopy. By applying a simple kinetic model, the small polaron formation time is found to increase from $70 \pm 10$ fs at 2.2 eV to $350 \pm 30$ fs at 2.6 eV. In comparison to a hematite thin film, the increased formation time can be explained by considering the differences in ligand field strength and Fe hopping center density that lead to altered electron hopping activation energies. Excitation energy-dependent analysis reveals a trend in polaron formation times that differs from that of hematite, but this trend may be due to a variety of differences between the samples. Further investigations, in particular a study of hematite nanoparticles, are required to separate the effects of crystallinity and morphology from the bonding and structural changes.

ASSOCIATED CONTENT

**Supporting Information**.

The following files are available free of charge.

Description of the XUV transient absorption experiment, sample fabrication and characterization, charge transfer multiplet modeling of the absorption spectra, description of the polaron formation kinetic model, and additional transient spectra with multivariate regression and kinetic model fits (PDF)

AUTHOR INFORMATION

**Corresponding Author**

*e-mail: srl@berkeley.edu

**ORCID**

Ilana J. Porter: 0000-0001-8692-9950 Scott K. Cushing: 0000-0003-3538-2259 Angela Lee: 0000-0001-5388-8400 Justin C. Ondry: 0000-0001-9113-3420 Hung-Tzu Chang: 0000-0001-7378-8212 A. Paul Alivisatos: 0000-0001-6895-9048 Stephen R. Leone: 0000-0003-1819-1338

**Notes**

The authors declare no competing financial interests.

ACKNOWLEDGMENT

This work was supported by the U.S. Department of Energy, Office of Science, Office of Basic Energy Sciences, Materials Sciences and Engineering Division, under Contract No. DEAC02-05-CH11231, within the Physical Chemistry of Inorganic Nanostructures Program (KC3103). S.K.C. acknowledges support by the Department of Energy, Office of Energy Efficiency and Renewable Energy (EERE) Postdoctoral Research Award under the EERE Solar Energy

Technologies Office. H.-T.C. acknowledges support by the Air Force Office of Scientific Research (AFOSR) (FA9550-15-1-0037). J.C.D. acknowledges support by the National Science Foundation Graduate Research Fellowship under DGE 1752814.